\documentclass{aastex}          %% The manuscript based on AASTeX v5.x

\usepackage{spr-astr-addons}    %% mimicing ASTR journal style
\usepackage{color}
\usepackage{rotating}

\newcommand{\D}{$^\circ$}

\newcommand{\HII}{H\,{\sc ii}}

\newcommand{\SII}{[S\,{\sc ii}]}

\newcommand{\OIII}{[O\,{\sc iii}]}

\newcommand{\HA}{H{$\alpha$}}

\newcommand{\Halpha}{H$\alpha$}

\newcommand{\nelc}{$n_e$}

\def\p0{\phantom{0}}

\def\arcmin{\hbox{$^\prime$}}
\def\arcsec{\hbox{$^{\prime\prime}$}}

\def\p0{\phantom{0}}
\def\apjl{ApJ}%
\def\apjs{ApJS}%
\def\ao{Appl.~Opt.}%
\def\aap{A\&A}%
\begin{document}

%% Article title

\title{Radio Planetary Nebulae in the Small Magellanic Cloud}

\shorttitle{RC PNe in the SMC}
\shortauthors{Leverenz}

\author{Howard Leverenz}
\affil{University of Southern Queensland, Toowoomba, Qld 4350, Australia}
\email{hleverenz@sbcglobal.net} %% non-output
\author{Miroslav D. Filipovi\'c}
\affil{University of Western Sydney, Locked Bag 1797, Penrith, NSW 2751, Australia}
\email{m.filipovic@uws.edu.au} %% non-output
\author{I. S. Boji\v ci\'c}
\affil{Department of Physics, The University of Hong Kong, Pokfulam Road, Hong Kong, China}
\author {E. J. Crawford}
\affil{University of Western Sydney, Locked Bag 1797, Penrith, NSW 2751, Australia}
\author{J. D. Collier}
\affil{University of Western Sydney, Locked Bag 1797, Penrith, NSW 2751, Australia}
\author{K. Grieve}
\affil{University of Western Sydney, Locked Bag 1797, Penrith, NSW 2751, Australia}
\author{D. Dra\v skovi\'c}
\affil{Department of Physics and Astronomy, Macquarie University, Sydney, NSW 2109, Australia}
\author{W. A. Reid}
\affil{Department of Physics and Astronomy, Macquarie University, Sydney, NSW 2109, Australia}

%\date{Accepted 2015 May XX. Received 2015 April XX; in original form 2015 April XX}

%\pagerange{\pageref{firstpage}--\pageref{lastpage}} \pubyear{2015}

\maketitle

%\label{firstpage}

\begin{abstract}
{
%\color{red}REDO
We present ten new radio continuum (RC) detections at catalogued planetary nebula (PN) positions in the Small Magellanic Cloud (SMC): SMP\,S6, LIN\,41, LIN\,142, SMP\,S13, SMP\,S14, SMP\,S16, J\,18, SMP\,S18, SMP\,S19 and SMP\,S22.  Additionally, six SMC radio PNe previously detected, LIN\,45, SMP\,S11, SMP\,S17, LIN\,321, LIN\,339 and SMP\,S24 are also investigated (re-observed) here making up a population of 16 radio detections of catalogued PNe in the SMC. These 16 radio detections represent $\sim$15 \% of the total catalogued PN population in the SMC. We show that six of these objects have characteristics that suggest that they are PN mimics: LIN\,41, LIN\,45, SMP\,S11, LIN\,142, LIN\,321 and LIN\,339. We also present our results for the surface brightness -- PN radius relation ($\Sigma$-D) of the SMC radio PN population. These are consistent with previous SMC and LMC PN measurements of the ($\Sigma$-D) relation.}
\end{abstract}


\section{Introduction}

 \label{s:intro}

The Small Magellanic Cloud (SMC) is a gas-rich late-type dwarf galaxy \citep{Bolatto2007} with a gas-to-dust ratio 30 times higher than the Milky Way \citep{Stanimirovic2000}. It is a member of the Local Group and is classified as an irregular galaxy \mbox{(ImIV-V)} \citep{Sandage1994}. The SMC is at a distance of $60.6 \pm 3.8$~kpc \citep{Hilditch2005} from the Galaxy which makes the spatial scale $\sim$0.3~pc/arcsec.

A planetary nebula (PN) is the result of an expanding envelope of gas that has been ejected near the end of the life of a star which has a zero age main sequence mass of $\sim$1 to 8~M$_{\odot}$ \citep{Pottasch1984}. Much of the gas is ejected as a stellar wind over an extended period of time during the Asymptotic Giant Branch (AGB) stage before it evolves into a Protoplanetary Nebula (PPN). When enough of the hydrogen envelope has been depleted from the PPN, the central star (CS) is exposed. The very hot, $\sim8\times10^4$~K, CS provides the energy to accelerate and ionize the gas forming a PN \citep{Pottasch1984,Bianchi1986}. This process has been quantified in the Interacting Stellar Winds Model in \cite{Kwok1978}.

\cite{Filipovic2009} reported the first confirmed extragalactic radio-continuum detection of 15 PNe in the Magellanic Clouds (MCs). Prior to this study, a tentative radio detection of only three extragalactic PNe had been reported in the literature \citep{1994A&A...290..228Z,2000A&A...363..717D}. Based on the radio-continuum properties of radio-bright Galactic PNe the expected radio flux densities at the distance of the SMC is up to $\sim$2.0~mJy at 1.4~GHz. However, \cite{Filipovic2009} found that two of the radio detected PNe in the SMC have flux densities  $>$4~mJy at 4.8~GHz, several times higher than NGC\,7027, the most radio-luminous known Galactic PN. This implied either a very high mass of the PN progenitor or the doubtful PN nature of these objects. In this paper we investigate these objects in the light of newly available radio and optical data.

The examination of PN evolutionary properties requires knowledge of parameters directly related to distances and total fluxes which are, in most cases, unknown or largely uncertain. However, since the distance to the MCs is known and highly refined, having the accurate PN radio-continuum flux densities can  provide an opportunity to accurately measure physical quantities such as electron density and ionised mass \citep{Crawford2012} which will be valuable in interpreting observations of PNe in our own Galaxy.

 \subsection{Radio-continuum emission from PNe}

Radio-continuum emission from PNe is primarily from free-free interactions between electrons and ions within the nebular shell \citep{1989agna.book.....O,Kwok2000}. The Spectral Energy Distribution (SED) is dependent on the metallicity, electron density and temperature. In the high frequency region, usually greater than 5GHz, the nebula is optically thin and the spectral index, $\alpha$\footnote{Where $\alpha$ is defined as $S \propto \nu^{\alpha}$ {with $S$ as the integrated flux density and $\nu$ as observed frequency.}}, is expected to be approximately $-0.1$. At lower frequencies, less than 5~GHz, young PNe could be optically thick with an expected spectral index of 2 in the approximation of a constant density nebula \citep{Pottasch1984}. For the case where the density profile is from a constant mass loss, the spectral index is expected to approach 0.6 \citep{Panagia1975}. However, \cite{Gruenwald2007} point out that radio SED observations from PNe cannot uniquely determine the density profile of the PNe.

The critical frequency of a PN is defined as the transition frequency from optically thick with a positive spectral index to optically thin with a negative spectral index. Most PNe measured in the Galaxy have a critical frequency of less than 5~GHz \citep{Gruenwald2007}. The critical frequencies for the PNe depend on gas density and are seen to range from approximately 0.4 to greater than 15~GHz \citep{Gruenwald2007}. \cite{Aaquist1991} observed young compact PNe with critical frequencies of approximately 5~GHz to 28~GHz. These young PNe may have very high electron densities which makes the critical frequency much higher than 5~GHz \citep{Kwok1982,Pazderska2009}.

Mechanisms and predictions for non-thermal radio emission, as discussed by \cite{Dgani1998}, require fast winds from the CS of the PN to create an inner region where strong magnetic fields are interacting with the CS wind. {A non-thermal SED would be expected to have a spectral index $\alpha\ll-0.1$ \citep{GurzadyanGrigorA.1997}. \cite{Cohen2006} reported emissions from a very unusual AGB star with shocked interactions between a cold PN--like nebular shell and the hot, fast stellar wind from the AGB star. The source of most apparent non-thermal radio detections is more likely to be a Supernova Remnant or Active Galactic Nuclei coincident with the position of the PN.

Previous examinations of PNe in the SMC have used data with resolutions of $\sim16\arcsec$ to $30\arcsec$ and with sensitivities of $\sim0.5$ to 0.7~mJy~beam$^{-1}$ (see Table \ref{tbl:data}). The improvements CABB brings to the ATCA's (Australia Telescope Compact Array) sensitivity allows us to examine the PNe with almost an order of magnitude improvement in resolution and sensitivity. This has led to the radio detection of ten SMC PNe previously undetected at radio frequencies with $\sigma\geqslant3$.

%% Table 1
%
 \begin{table}
 \scriptsize
 \caption{ {Parameters of the radio data used in this study.} } %% no full stop at the end of caption
 \label{tbl:data}
 \begin{tabular}{lcccc}
 \hline  %% rule at top

 Radio   & Frequency & Sensitivity     & Resolution          & Ref \\
 Band    & (GHz)     &(mJy~beam$^{-1}$)& (\arcsec)           &           \\
 \hline
 % \tablenotemark{a}
 $3~cm^a$    & 8.640      & 0.6             & 20                  & 1 \\
 $3~cm^b$    & 9.000      & 0.027           & 1.27 $\times$ 0.97  & 2 \\
 $6~cm^a$    & 4.800      & 0.7             & 30                  & 1 \\
 $6~cm^b$    & 5.500      & 0.022           & 2.06 $\times$ 1.59  & 2 \\
 $6~cm^c$    & 5.466      & 0.07            & 2.74 $\times$ 2.22  & 3  \\
 $13~cm^a$   & 2.100      & 0.015           & 1.7                 & 4 \\
 $13~cm^b$   & 2.100      & 0.06            & 1.7                 & 5 \\
 $20~cm^a$   & 1.420      & 0.5             & 16.33 $\times$ 9.45 & 6  \\
 $20~cm^b$   & 1.420      & 0.1             & 5.3 $\times$ 5      & 7 \\
 $20~cm^c$   & 1.370      & 0.05            & 7.05 $\times$ 6.58  & 8 \\
 \hline
 \end{tabular}
 \smallskip
 \flushleft
 References: {1:}\cite{Crawford2011}, {2:} \cite{Crawford2012} CABB, {3:} ATCA C2768 (PI: Jordan Collier) CABB, {4:} ATCA C2521 (PI: Jacco van Loon) CABB, {5:} ATCA CX310 (PI: Andrew O'Brian) CABB, {6:} \cite{Wong2011}, {7:} \cite{Wong2012a}, {8:} \cite{Bojicic2010}\\
 \end{table}

\begin{figure*}
 \begin{center}
 \includegraphics[angle=-90, trim=0 50 0 0, width=1.175\textwidth]{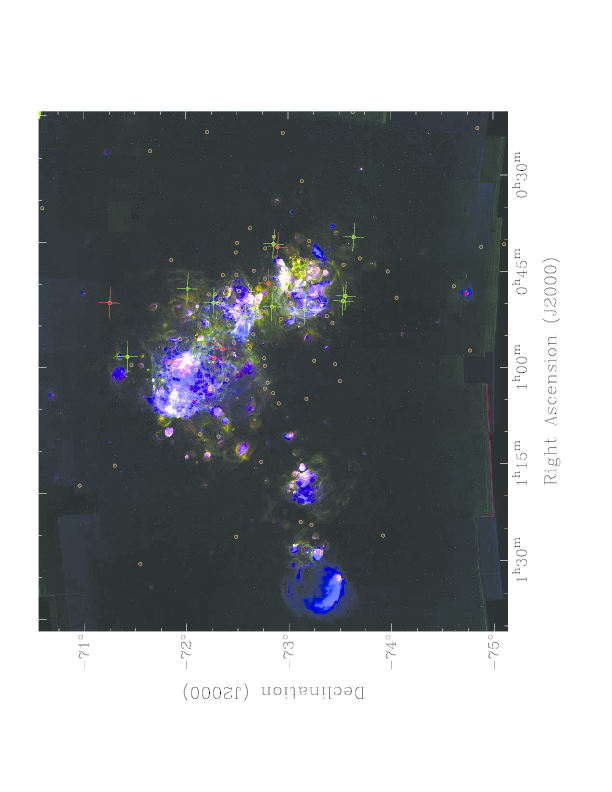}
 \caption{Distribution of the 105 catalogued PNe in the SMC examined here (small orange circles). The background SMC image \citep[MCELS:][]{Smith2000} consists of H$\alpha$ (red), \SII\ (green) and \OIII\ (blue) images. The ten locations highlighted with the green crosses are at the coordinates of the new radio PNe. The six red crosses are locations of previously detected radio PNe.}
 \label{fig:PNeSMC}
 \end{center}
\end{figure*}

\section{Archival data and new radio-continuum observations}
 \label{s:data}

\begin{figure*}[t]
\centering
\includegraphics[trim=0 0 0 0, width=.32\textwidth]{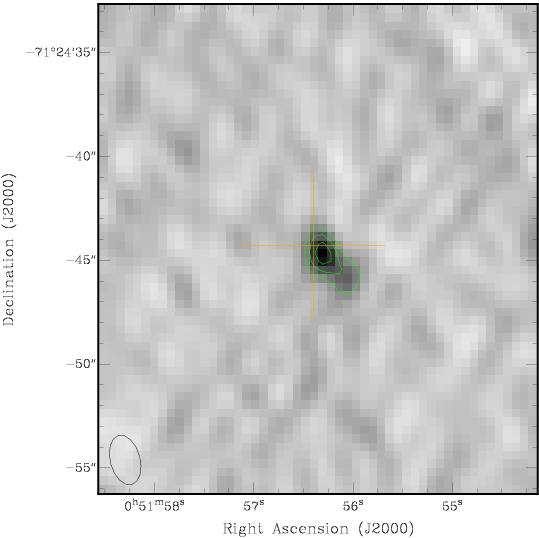}
\includegraphics[trim=0 10 0 -10, width=0.33\textwidth]{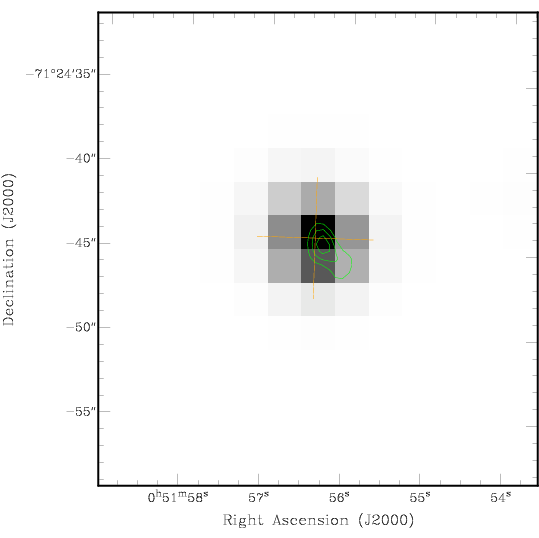}
\includegraphics[trim=0 0 0 0, width=0.32\textwidth]{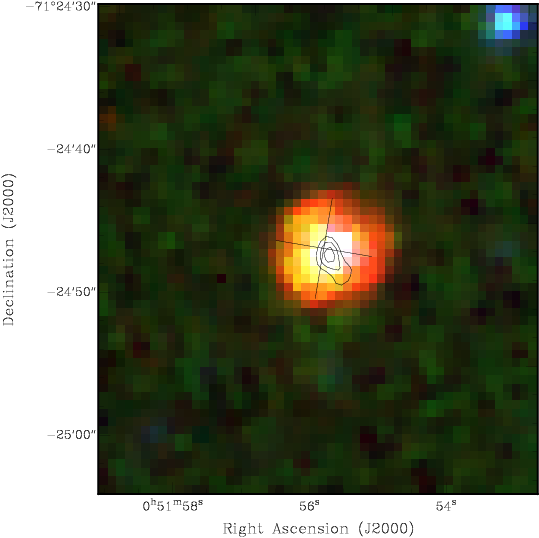}
\caption{Example of finding charts for SMC PNe. Images are constructed as: {\it Left}: total intensity radio image, {\it Middle}: \HA\ cutout image from MCELS data and {\it Right}: colour composite images from the IRAC (GLIMPSE-SMC) data with RGB = 8.0, 5.8 and 4.5~$\micron$. All images are overlaid with radio contours (green/black full lines) and catalogued SIMBAD position (orange/black cross). The resolving beam in the radio image is shown in the lower left corner. Field sizes are approximately the same. The object shown in this example is SMP\,S17.}
\label{fig:findingcharts}
\end{figure*}

In the search for SMC PN radio counterparts we used several archival ATCA datasets together with a targeted observation of a sample of SMC PNe.  Table~\ref{tbl:data} presents compiled information on radio-continuum imagery used in this study.

\cite{2011SerAJ.182...43W} described the creation of a new high resolution image at 20~cm for the SMC. This was done by combining three archived images from ATCA projects (C159, C1288 and C1197) plus the Parkes image from \cite{Filipovic1997}. The images included array configurations of 375-m, 6A, H214 and H75. This is the $20~cm^a$ image.

\cite{Wong2012a} studied compact \HII\ regions in the SMC. Archive images were used at 3, 6, 13, 20 and 36~cm \citep{Wong2011a,Filipovic2002,Turtle1998} plus images from the area near LHA 115-N~77 from \cite{Ye1995}. Additional images from near LHA 115-N~19 were used including ATCA images from array configurations of 375-m, 1.5D, 6C, 6B and 750A. The 20\,cm image produced by their study and used in our paper is referred to as the $20~cm^b$ image.

\cite{Bojicic2010} created a new high resolution image from ATCA project C281 (this data was originally acquired by \cite{Ye1995}). Their reprocessing of the C281 data using a different cleaning technique and careful flagging of the noisy observational data resulted in the new high resolution, high sensitivity image used here.  This image data set is referred to in our paper as the $20~cm^c$ image.

The $13~cm^a$ ($\nu$=2.1~GHz) image used in this study is from ATCA project C2521 (PI: Jacco van Loon). These observations were obtained with the CABB in January 2012 with the ATCA configured in the 6A array. The $13~cm^b$ ($\nu$=2.1~GHz) image used here is from our new, December 2014, CABB observations (Project CX310) (PI: Andrew O'Brian) with the ATCA configured in the 6A array.

The $6~cm^c$ image used in this study is from ATCA project C2768 (PI: Jordan Collier; observation from 6$^{th}$ February 2013, J. D. Collier et al. 2016, in prep.) obtained with the CABB. This image was acquired with the ATCA 6A antenna configuration.

\cite{2011SerAJ.183...95C} applied a ``peeling'' technique employing joint deconvolution to 6 and 3~cm images from the SMC. Images from five ATCA projects (C1604, C1207, C882, C859, C634) were combined. These are referred to as the $3~cm^a$ and $6~cm^a$ images.

Our new ATCA observations, project C2367 (PI: I.~Boji\v ci\'c) aimed to unravel the bright end of the radio PN luminosity function (PNLF) from new radio samples in Magellanic Clouds, understand the quantitative differences in the multi-wavelength characteristics of Galactic and MCs PNe and relate these to PN age and luminosity (Boji\v ci\'c et al. in preparation). The LMC PN sample contains observations of 40 objects populating the bright end  (first four magnitudes)  of the \OIII~PNLF. In the SMC observations we sample only 10 objects in  a larger range of  \OIII\ magnitudes in order to determine their general radio properties and resolve some of the dubious detections and/or PN identifications. The SMC data were acquired on $28^{th}$ and $29^{th}$ of November 2010 using the ``snapshot'' mode with $\sim1$~hour integration for each pointing over the 12~hour observing session. We processed the images using {\sc miriad} \citep{Sault1995}, {\sc karma} (Gooch, R., Barnes, J.: 1996 Karma: a Visualisation Testbed) and \cite{Scilab2013} software packages. These are referred to as the $3~cm^b$ and $6~cm^b$ images.

\begin{table*}[ht]
\centering
\scriptsize
\caption{Coordinates and flux density measurements of detected objects.  The new flux density measurements are shown in the columns titled ``new'' and previous detections from  \cite{Payne2008,Filipovic2009,Bojicic2010} are listed in the columns titled ``F09''. A colon in front of a radio flux density represents an uncertain radio detection. A ``$>$'' indicates that the number is the lower limit for a flux density. Superscripts {\it a}, {\it b}, and {\it c} refer to specific data sources referenced in the ``Radio Band'' column of Table~\ref{tbl:data}}.  \\
 \begin{tabular}{rlcccccccccl}
 \hline
No.	&PN			&RAJ2000	&DECJ2000		&\multicolumn{2}{c}{S$_{3cm}$~(mJy)}&\multicolumn{2}{c}{S$_{6cm}$~(mJy)}&\multicolumn{2}{c}{S$_{13cm}$~(mJy)}&	\multicolumn{2}{c}{S$_{20cm}$~(mJy)}\\
	& 			&			&				&new			&F09&new			&F09	&new			&F09	&new				&F09\\
     	&(1)  		&(2)			&(3)				&(4)			& (5)	&(6)			&(7)		&(8)			&(9)		&(10)			&(11)\\
 \hline
1	&SMP\,S6	&00:41:27.8	&--73:47:06		&1.1$\pm0.5^b$&...	&1.3$\pm0.1^b$&...		&...			&...		&$>$0.2$^b$		&...\\
2	&LIN\,41		&00:43:12.9	&--72:59:58		&...			&...	&...			&...		&...			&...		&0.6$\pm0.2^b$	&...\\
3	&LIN\,45		&00:43:36.7	&--73:02:27		&...			&3.1	&4.8$\pm0.3^b$&4.4		&...			&5.0		&4.4$\pm0.5^b$	&5.1\\
4	&SMP\,S11	&00:48:36.5	&--72:58:01		&...			&2.4	&2.9$\pm0.5^a$&2.6		&...			&2.0		&2.0$\pm0.2^b$	&1.9\\
\smallskip
5	&LIN\,142		&00:49:46.4	&--73:10:26		&...			&...	&...			&...		&:0.32$\pm0.12^a$&...	&...				&...\\
6	&SMP\,S13   &00:49:51.6	&--73:44:23		&...			&...	&...			&...		&0.6$\pm0.2^a$&...		&...				&...\\
7	&SMP\,S14	&00:50:35.1	&--73:42:58		&...			&...	&...			&...		&$0.4\pm0.1^a$&...		&...				&...\\
8	&SMP\,S16	&00:51:27.2	&--72:26:11		&...			&...	&0.5$\pm0.2^c$&...		&...			&...		&...				&...\\
9	&J\,18		&00:51:43.7	&--73:00:53		&...			&...	&...			&...		&:0.24$^a$	&...		&...				&...\\
\smallskip
10	&SMP\,S17   &00:51:56.3	&--71:24:45		&...			&...	&0.9$\pm0.3^b$&1.5		&...			&...		&:1.0$^a$			&1.4\\
11	&SMP\,S18	&00:51:58.3	&--73:20:32		&0.9$\pm0.5^b$&...	&0.8$\pm0.15^b$&...		&...			&...		&...				&...\\
12	&SMP\,S19	&00:53:11.1	&--72:45:08		&...			&...	&...			&...		&0.6$\pm0.2^b$&...		&...				&...\\
13	&LIN\,321		&00:57:29.9	&--72:32:24		&:1.5$^a$		&...	&3.5$\pm1.5^a$		&3.0$^\ddag$	&...		&...		&4.4$\pm1.5^a$	&...\\
14	&SMP\,S22	&00:58:37.1	&--71:35:50		&...			&...	&0.4$\pm0.2^b$&...		&...			&...		&...				&...\\
\smallskip
15	&LIN\,339		&00:58:42.9	&--72:27:17		&:1.3$^a$		&4.2	&4.4$\pm0.9^a$&4.1		&...			&3.5		&$>$1.1$^c$		&2.3\\
16	&SMP\,S24	&00:59:16.1	&--72:02:00		&:0.5	$^b$		&...	&0.7$\pm0.2^b$&...		&...			&...		&...				&0.7$^\dag$\\
  \hline
\end{tabular}
 \smallskip
  \flushleft
$^\ddag$ \cite{Payne2008};$^\dag$ \cite{Bojicic2010}
\label{tbl:MeasureRC}
\end{table*}

\section{Radio Detections of PNe in the SMC}
 \label{s:RC Detection}

\subsection{Base catalogue and detection method}

In order to make the base catalogue for this study we used the SIMBAD database \citep{Wenger2000}. The SIMBAD database  was queried for objects with PN as the primary object type and within the approximate boundaries of the 20~cm image from \cite{Wong2011} which extends approximately 3\D\ from the centre of the SMC: RA(J2000)=00$^h$52$^m$38$^s$ and Dec(J2000)=--72\D48\arcmin01\arcsec (see Fig.~\ref{fig:PNeSMC}). Within the 3\D\ boundary, 105 catalogued SMC PNe were found. We also searched the literature for any other PNe or PN candidates not listed in SIMBAD. However, no other objects which fall into this category were found. Since the examined radio images each have different coverage of the SMC, each of the catalogued PN locations were examined to see if they were contained within the image boundaries of any of the images listed in Table~\ref{tbl:data}. For those cases where an image contained the PN coordinates, the image was examined for PN radio emissions.

To confirm the radio detection we examined the H$\alpha$ cutout images from {\it Magellanic Cloud Emission-line Survey} \citep[MCELS:][]{Smith2000} and MIR images from {\it Surveying the Agents of Galaxy Evolution in the Tidally Stripped, Low Metallicity Small Magellanic Cloud} (SAGE-SMC) Spitzer Legacy program \citep{Gordon2011} overlaid with contours from the corresponding radio map for each PN. As positive detections we considered objects with peak flux over 3~$\sigma$, where $\sigma$ is the measured noise level. We adopted the positional uncertainty as the limit of the FWHM fits to the data convolved with the ATCA maximum absolute positional uncertainty of 5\arcsec\ \citep{Stevens2014}. We also required that the distance from the catalogued position to the radio peak be within the positional uncertainties of the radio peak. The measured positional uncertainties for all of the objects presented in this paper meet this requirement. We present an example of finding charts in Fig.~\ref{fig:findingcharts}.

 \subsection{Detection Results} \label{sec:detresults}

After careful examination of all available data we found ten new radio-continuum detections of objects in our base catalogue: SMP\,S6, LIN\,41, LIN\,142, SMP\,S13, SMP\,S14, SMP\,S16, J\,18, SMP\,S18, SMP\,S19 and SMP\,S22. Also, for six previously detected objects \citep{Payne2008,Filipovic2009, Bojicic2010}: LIN\,45, SMP\,S11, SMP\,S17, LIN\,321, LIN\,339 and SMP\,S24, we recorded measurement of flux density in at least one additional frequency, confirming the previous detection.

\begin{figure}[h]
\centering
\includegraphics[width=0.33\textwidth,angle=-90]{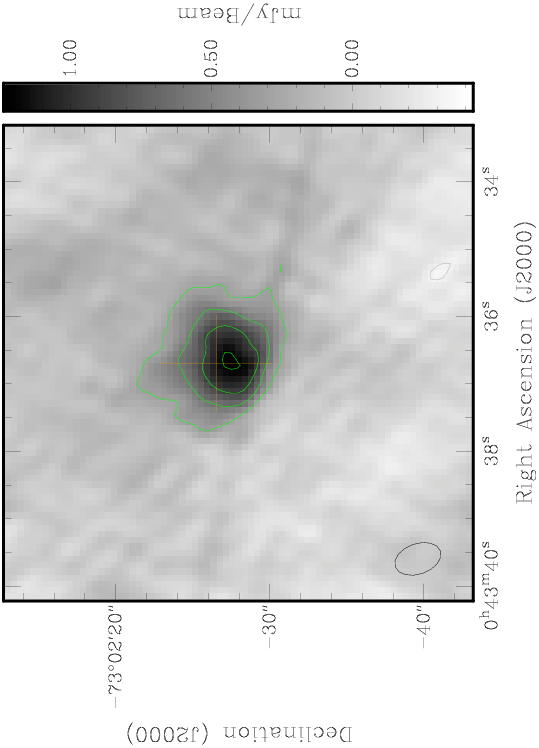}\
\caption{The radio-continuum image at 6~cm of LIN\,45 overlaid with contours at 3, 6, 9 and $12\times\sigma$ where $\sigma$=0.1~mJy/beam is the RMS noise measured in the vicinity of the detection. The resolving beam is shown in the lower left corner. The orange cross represents the catalogued position for this object. Note that this object is clearly resolved at 3 and $6\times\sigma$ with an angular extension of about 7\arcsec\ (the ellipse at lower left corner is image beam size).}
\label{fig:lin456cm}
\end{figure}

In targeted observations of SMC PNe (images $3~cm^b$ and $6~cm^b$) we detected 6 out of 10 observed objects. We did not find radio counterparts for:  SMP\,S10, SMP\,S25, SMP\,S7 or SMP\,S9.

Except for one (LIN~45 at 6~cm; see Fig.~\ref{fig:lin456cm}) all detected radio counterparts appear to be unresolved in our radio maps. Therefore, in order to estimate integrated flux densities and positions, the radio image at each PN position was fitted with a Gaussian model using the {\sc miriad imfit} task. The noise for each radio map was estimated by applying the {\sc miriad sigest} task to an empty part of the image (i.e. with no emission or point sources), with a minimum area of 10~arcmin$^{2}$.
%Integrated flux for LIN~45 at 6~cm is estimated using an aperture photometry method (see for example {\color{red} Miro cite some of your snr papers...}).

The SIMBAD coordinates and the measured integrated radio fluxes for detected objects are presented in Table~\ref{tbl:MeasureRC}. Since none of the detected objects appear to be resolved we determined uncertainties in flux density estimates from the noise (RMS). We present finding charts for newly detected SMC PNe (excluding the six objects previously detected) in Figs.~\ref{fig:radio_gr1a},~\ref{fig:radio_gr1b}~and~\ref{fig:radio_gr1c}. These images are constructed as total intensity radio maps with at least one of the detection frequencies overlaid with contours at multiples of the estimated RMS noise. Orange crosses represent previously catalogued positions found in SIMBAD.

\begin{figure*}%[H]
\centering
\includegraphics[trim=0 0 0 0, width=.32\textwidth]{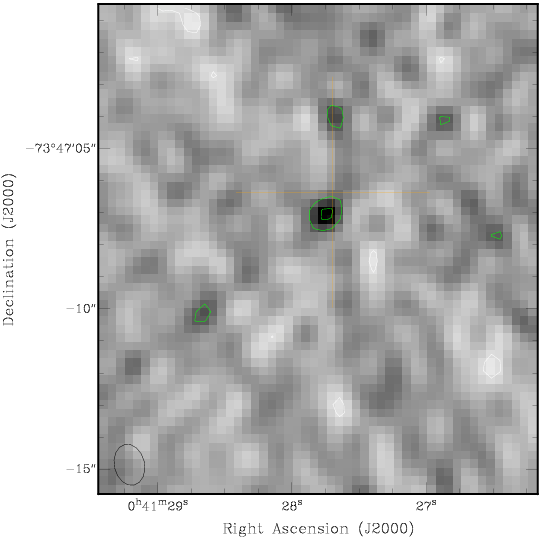}
\includegraphics[trim=0 0 0 0, width=.32\textwidth]{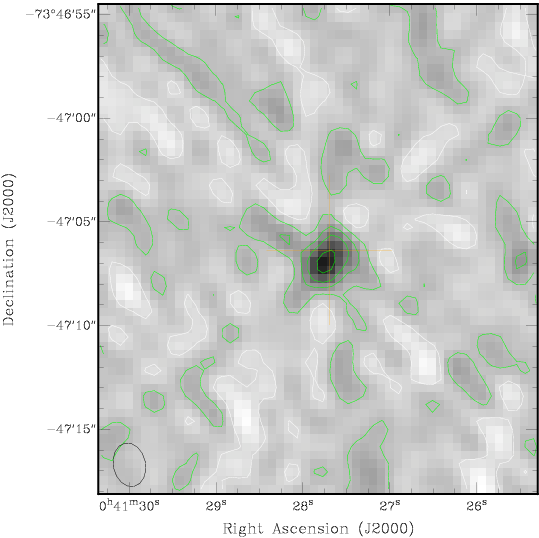}
\includegraphics[trim=0 0 0 0, width=.32\textwidth]{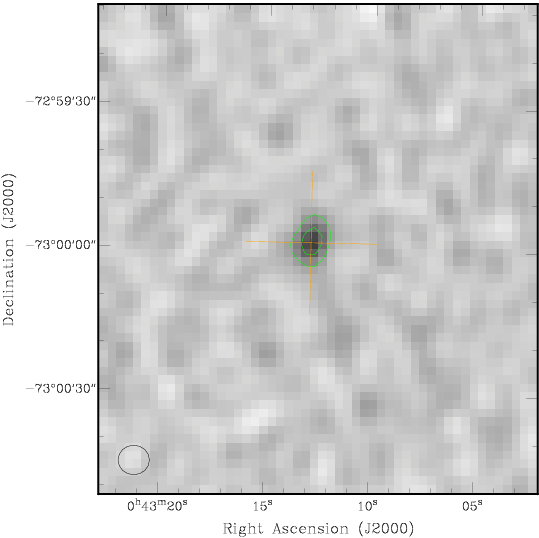}
\includegraphics[trim=0 0 0 0, width=.32\textwidth]{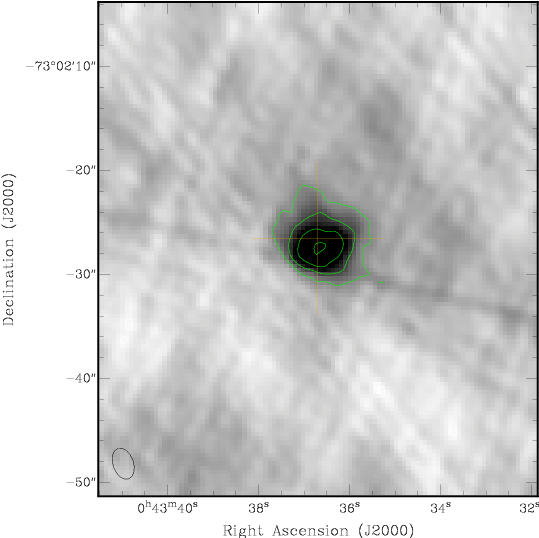}
\includegraphics[trim=0 0 0 0, width=.32\textwidth]{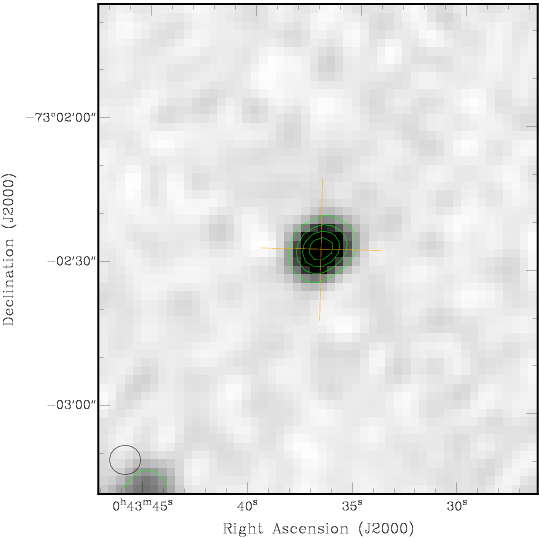}
\includegraphics[trim=0 0 0 0, width=.32\textwidth]{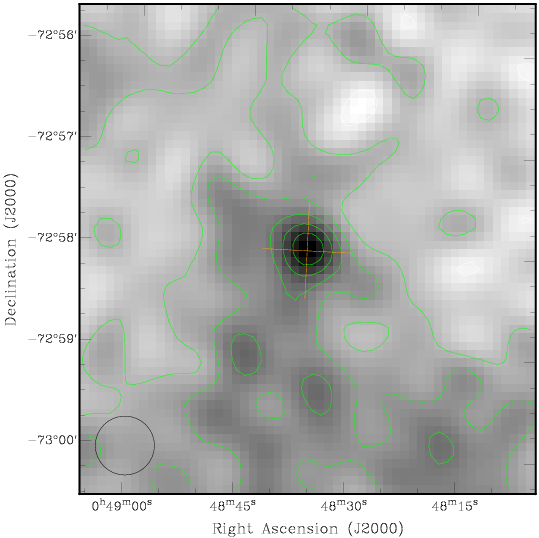}
\includegraphics[trim=0 0 0 0, width=.32\textwidth]{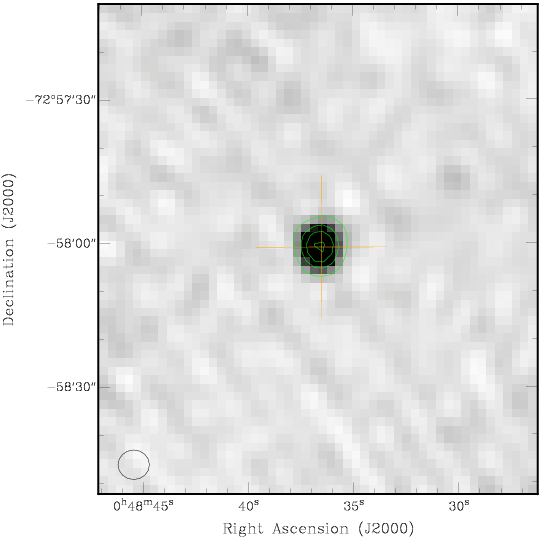}
\includegraphics[trim=0 0 0 0, width=.32\textwidth]{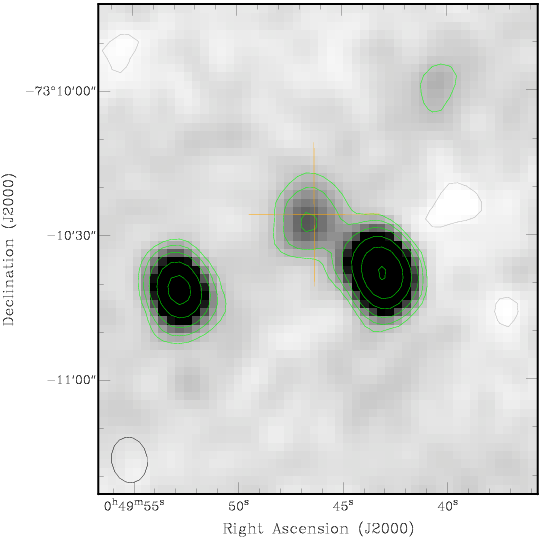}
\includegraphics[trim=0 0 0 0, width=.32\textwidth]{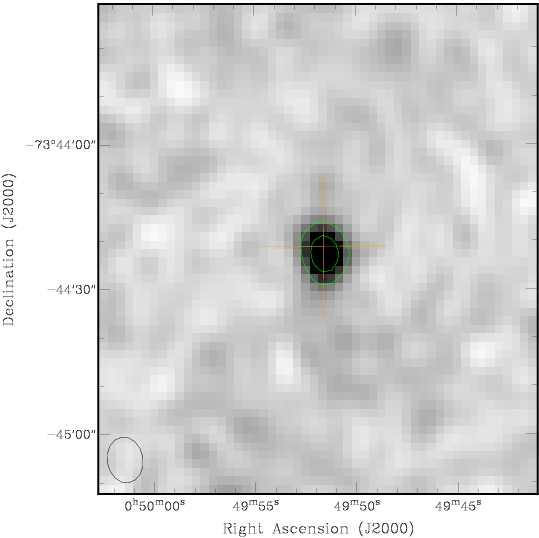}
\caption{New radio detection of SMC PNe. Images are constructed as total intensity radio maps. Contours are integral multiples of the measured RMS noise. Images are left to right, top to bottom: SMP\,S6~(at 3~cm), SMP\,S6~(at 6~cm), LIN\,41~(at 20~cm), LIN\,45~(at 6~cm), LIN\,45~(at 20~cm), SMP\,S11~(at 6~cm), SMP\,S11~(at 20~cm), LIN\,142~(at 13~cm), SMP\,S13~(at 13~cm). The beam size of each image is shown in the bottom left corner.}
\label{fig:radio_gr1a}
\end{figure*}

\begin{figure*}%[H]
\centering
\includegraphics[trim=0 0 0 0, width=.32\textwidth]{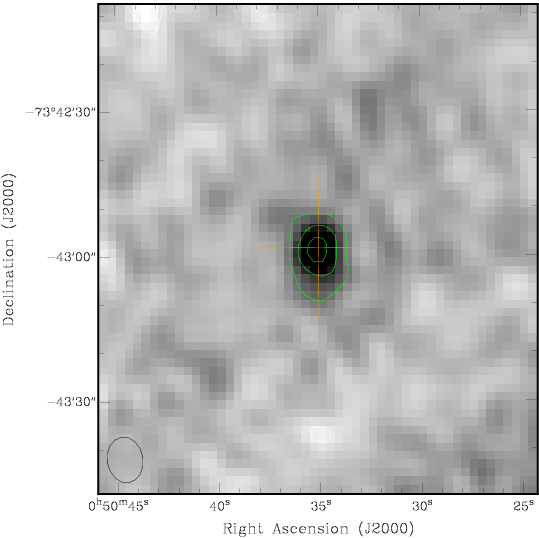}
\includegraphics[trim=0 0 0 0, width=.32\textwidth]{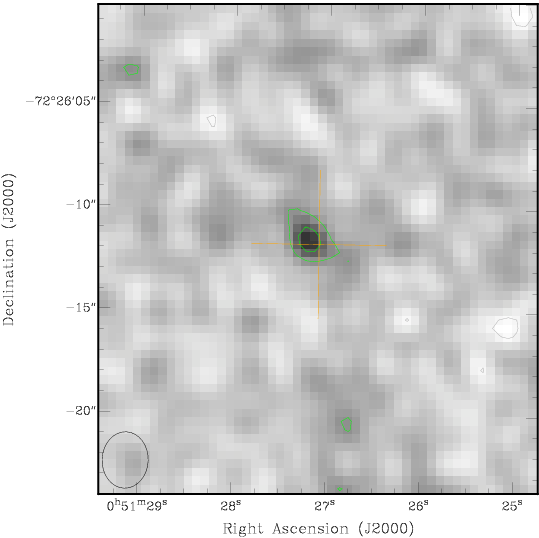}
\includegraphics[trim=0 0 0 0, width=.32\textwidth]{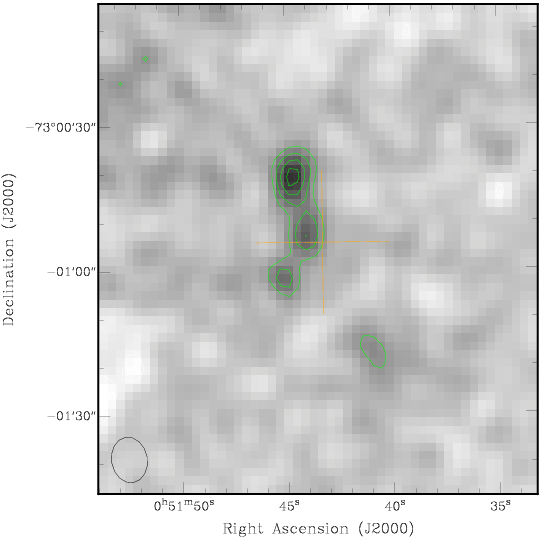}
\includegraphics[trim=0 0 0 0, width=.32\textwidth]{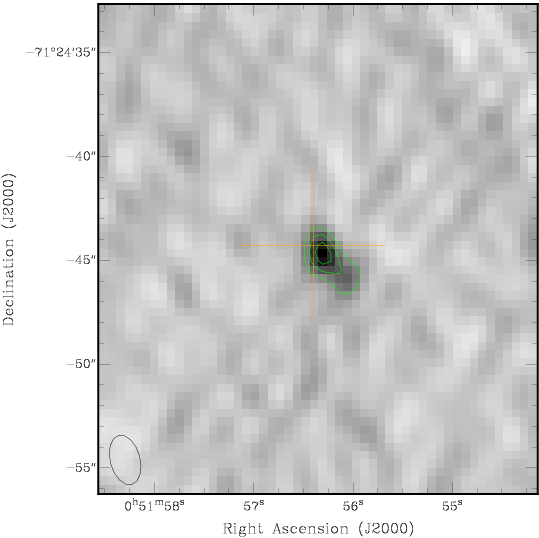}
\includegraphics[trim=0 0 0 0, width=.32\textwidth]{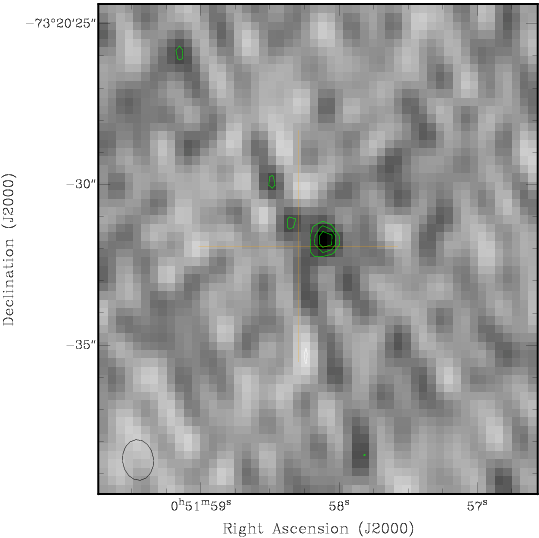}
\includegraphics[trim=0 0 0 0, width=.32\textwidth]{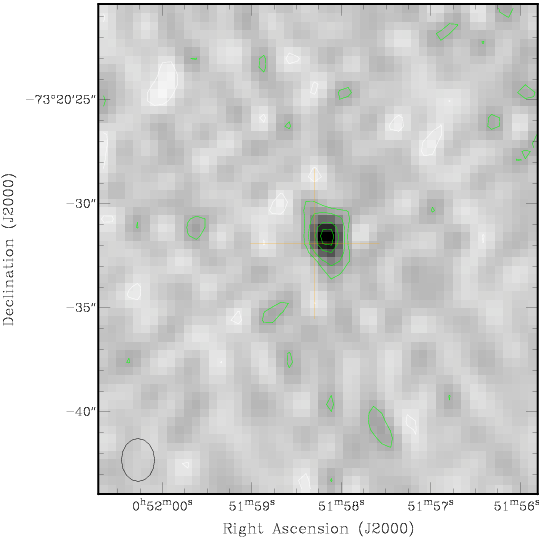}
\includegraphics[trim=0 0 0 0, width=.32\textwidth]{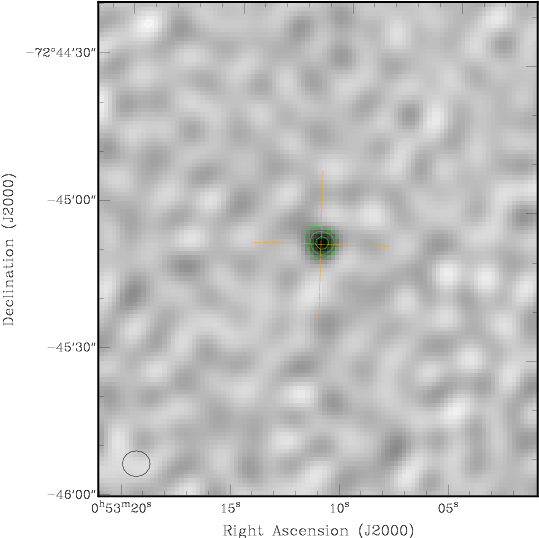}
\includegraphics[trim=0 0 0 0, width=.32\textwidth]{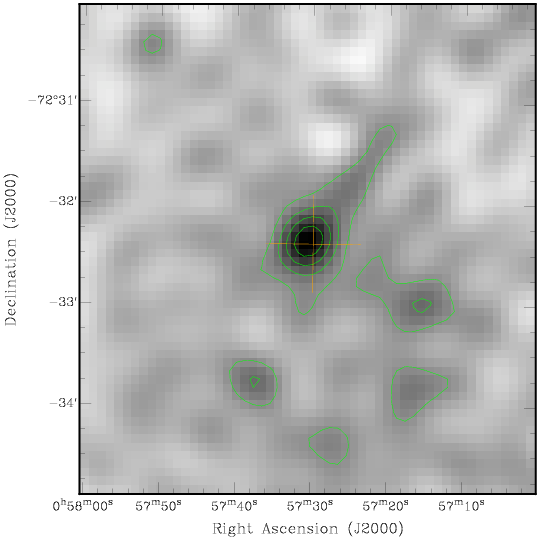}
\includegraphics[trim=0 0 0 0, width=.32\textwidth]{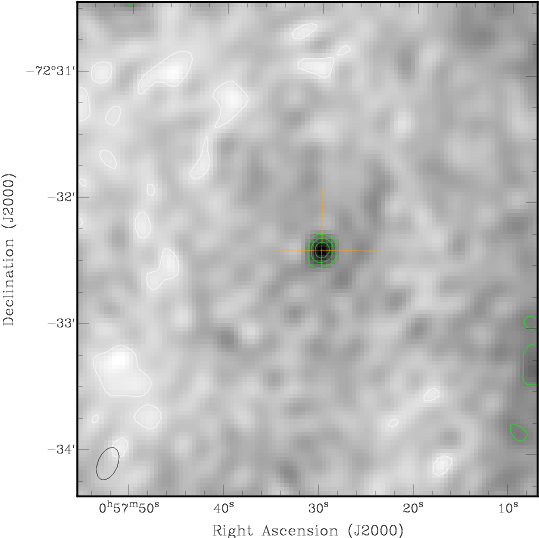}
\caption{New radio detection of SMC PNe. Images are constructed as total intensity radio maps. Contours are integral multiples of the measured RMS noise. Images are left to right, top to bottom: SMP\,S14~(at 13~cm) SMP\,S16~(at 6~cm), J\,18~(at 13~cm), SMP\,S17~(at 6~cm), SMP\,S18~(at 3~cm), SMP\,S18~(at 6~cm), SMP\,S19~(at 13~cm), LIN\,321~(at 6~cm), LIN\,321~(at 20~cm). The beam size of each image is shown in the bottom left corner.}
\label{fig:radio_gr1b}
\end{figure*}

\begin{figure*}%[H]
\centering
\includegraphics[trim=0 0 0 0, width=.32\textwidth]{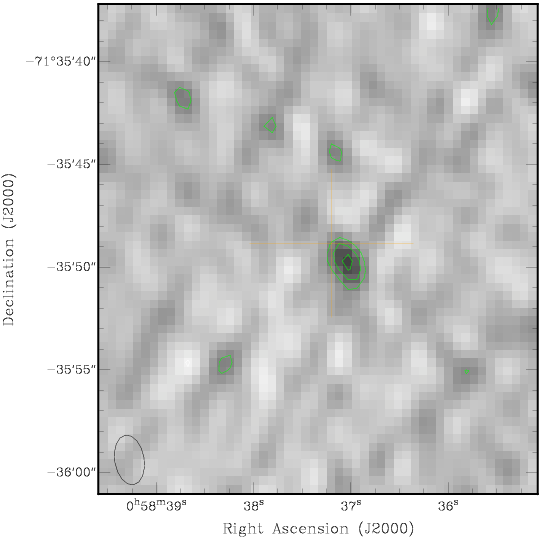}
\includegraphics[trim=0 0 0 0, width=.32\textwidth]{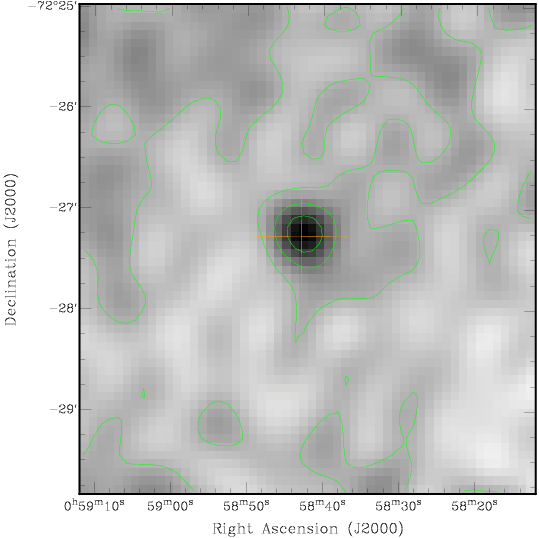}
\includegraphics[trim=-30 0 30 0, width=.32\textwidth]{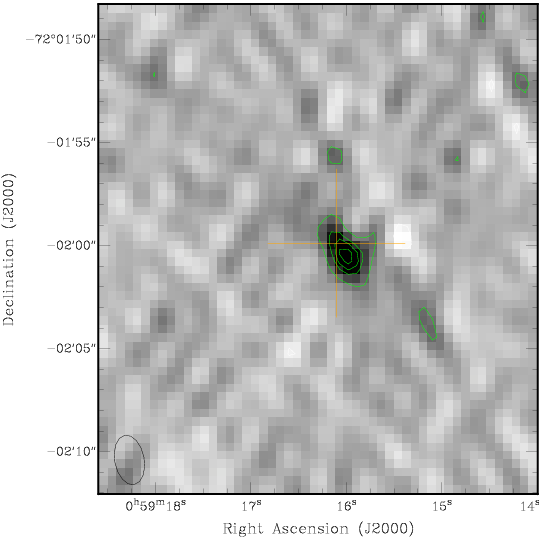}
\caption{New radio detection of SMC PNe. Images are constructed as total intensity radio maps. Contours are integral multiples of the measured RMS noise. Images are left to right, top to bottom: SMP\,S22~(at 6~cm), LIN\,339~(at 6~cm), SMP\,S24~(at 6~cm). The beam size of each image is shown in the bottom left corner.}
\label{fig:radio_gr1c}
\end{figure*}

As can be seen from the finding charts presented, in most cases the radio detections are of isolated sources well correlated with their optical counterparts. Only one object (J\,18) show a relatively large offset ($\sim$4\arcsec) and appears to be associated with a large and complex radio structure. In combination with the overall brightness in optical and infrared bands of J\,18, compared to other radio detected objects, and this unusual offset, we decided to flag this detection only as possible.

\begin{table*}
\centering
\scriptsize
\caption{Measured and calculated SMC RC PN parameters. The photometric angular diameters (Col.~2)  are from the sources referenced in Col.~3.  The spectral indices ($\alpha$) are shown in Col.~4.  The adopted flux densities at 6~cm are shown in Col.~5. The radii in pc are shown in Col.~7. Values for brightness temperatures (T$_b$), the electron densities (n$_e$) and ionised masses (M$_i$) are calculated from the adopted 6~cm fluxes and presented in Cols.~(6), (8) and (9), respectively. } % %% no full stop at the end of caption
 \begin{tabular}{lccrcrccccccc}
 \hline
 \hline
PN			&$\theta$	&ref$_\theta$&$\alpha$&S$_{6cm}$		&T$_b$	&r		&log \nelc			&$M_i$\\
			&('')		&		&		&(mJy)			&(K)		&(pc)	&($cm^{-3}$)				&($M_{\odot}$)\\
(1)			&(2)		&(3)		&(4)		&(5)				&(6)		&(7)		&(8)				&(9)\\
  \hline
SMP\,S6		&0.4		&2		&--0.3	&1.3$\pm$0.1		&620			&0.057	&4.2			&0.12\\
LIN\,41		&...		&...		&...		&0.5$\pm$0.2		&...			&...		&...			&...\\
LIN\,45		&9.0		&1		&--0.15	&4.6$\pm$0.4		&$<$10		&2.7		&2.4			&22.30\\
SMP\,S11		&0.7$^w$	&2		&0.16	&2.7$\pm$0.4		&420			&0.105	&3.2			&1.54\\
\smallskip
LIN\,142		&14		&1		&...		&:0.29			&$<$10		&4.2		&1.5			&10.56\\
SMP\,S13		&0.4		&2		&...		&0.6$\pm$0.2		&290			&0.057	&4.0			&0.08\\
SMP\,S14		&0.9		&2		&...		&0.35$\pm$0.15	&40			&0.126	&3.4			&0.17\\
SMP\,S16		&0.4		&3		&...		&0.5$\pm$0.2		&240			&0.054	&3.9			&0.06\\
J\,18		&0.4		&2		&...		&:0.22			&:110		&0.051	&:3.8			&:0.05\\
\smallskip
SMP\,S17		&0.5		&2		&--0.13	&1.2$\pm$0.3		&370			&0.075	&4.0			&0.15\\
SMP\,S18		&0.3		&2		&0.19	&0.80$\pm$0.15	&680			&0.045	&4.2			&0.05\\\
SMP\,S19		&0.6		&2		&...		&0.6$\pm$0.2		&130			&0.09	&3.6			&0.10\\\
LIN\,321		&3.3		&4		&--0.18	&3.5$\pm$1.5		&30			&0.5		&3.0			&4.37\\
SMP\,S22		&0.8		&2		&...		&0.4$\pm$0.2		&50			&0.12	&3.5			&0.20\\
\smallskip
LIN\,339		&2.7		&4		&0.26	&4.3$\pm$1.1		&50			&0.4		&3.2			&3.80\\
SMP\,S24		&0.4		&2		&0		&0.7$\pm$0.2		&350			&0.06	&4.0			&0.08\\
  \hline
\end{tabular}
 \smallskip
 \flushleft
1:  \cite{Bica2000} 2: \cite{Stanghellini2003}, 3: \cite{Meynadier2007}, 4: \cite{Shaw2010}
 \label{tbl:physprop}
\end{table*}

\subsection{Radio-continuum properties of the detected sample} \label{sec:radprop}

In order to make additional estimates of physical properties from the objects in our sample we investigated the RC spectral energy distribution and brightness temperature (T$_b$) for each object as good proxies for PN evolutionary status  \citep{Kwok1985,zijlstra1990,Gruenwald2007}. The estimated spectral indices for objects having measurements of flux density on at least two frequencies are presented in Table~\ref{tbl:physprop}. Two objects from our sample show possible self-absorption effects at lower frequencies (SMP\,S11 and LIN\,339). We further examined the SEDs of these two objects by fitting radio SED models from \cite{Siodmiak2001} with assumed spherical symmetry and a constant density distribution. The results are shown in Fig.~\ref{fig:compars} (left). Although there is an indication that the emission in the $\nu<1.5$~GHz could be affected by self-absorption, the emission at 5~GHz appears to be optically thin.  We conclude that none of the investigated objects show optical thickness effects in the high frequency region ($\nu>2$~GHz) and the measurements are good to use for further calculation of physical parameters. Column~5 in Table~\ref{tbl:physprop} presents the adopted flux densities at 5~GHz for all of the detected objects. If direct measurement at 5~GHz was not available, we estimated the flux density from a neighbouring band assuming optically thin emission ($\alpha=-0.1)$.

\begin{figure*}[ht]
\centering
\includegraphics[trim=0 0 0 0, width=.48\textwidth]{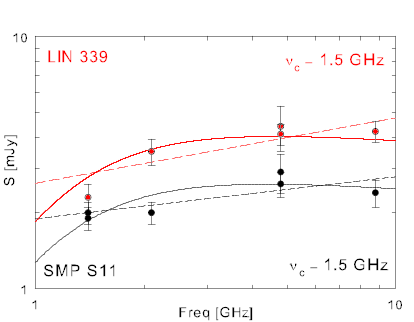}
\includegraphics[trim=0 0 0 0, width=.5\textwidth]{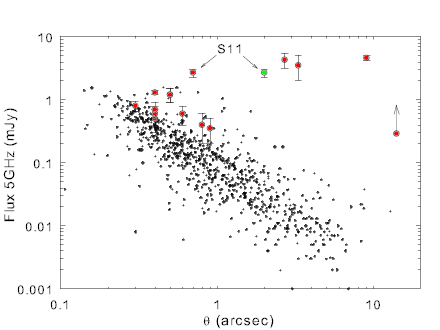}
\caption{{\it Left}: Radio SED model fits from \cite{Siodmiak2001} of LIN\,339 and SMP\,S11. {\it Right}: Galactic PNe data from Boji\v ci\'c 2010 (PhD Thesis) and \cite{frew2015} scaled to the SMC distance (black dots) and our new SMC measurements (red markers).}
\label{fig:compars}
\end{figure*}

Finally, for all of our objects we calculated T$_b$ at 5~GHz using available, optically determined, angular diameters (Table~\ref{tbl:physprop}; Col.~2). The angular diameters were adopted from \cite{Bica2000,Stanghellini2003,Meynadier2007,Shaw2010} and presented in Table~\ref{tbl:physprop}. From \cite{Stanghellini2003} we used photometric radii except for SMP~S11 where the measured angular dimension is at 10\% of peak brightness. Radio brightness temperatures for this object at 1.4~GHz, calculated from $\theta=2$\arcsec\ and $\theta=0.7$\arcsec, were 350~K and 3700~K, respectively. Since the former T$_b$ is in better agreement with the indicated self absorption effects at $\nu<1.5$~GHz we adopted $\theta=0.7$\arcsec\ as the better estimate of this object's angular size. Low brightness temperatures for all objects confirm optically thin emission in the observed frequency range.

We compared the obtained results with a sample of known Galactic PNe (GPNe). Recently, \cite{frew2015} published well determined distances for over 1200 GPNe. We used these results to scale angular diameters and integrated flux densities for $\sim$750 GPNe to the distance of the SMC. The radio flux densities were adopted from Boji\v ci\'c 2010 (PhD Thesis) and references therein.  In Fig.~\ref{fig:compars} (right) we present the distribution of flux densities vs angular diameters ($S-\theta$) for a sample of GPNe (black crosshairs) and radio detected SMC PNe (red filled circles; excluding J\,18 because of its doubtful detection and LIN\,41 because no published angular diameter is available). An additional data point for SMP\,S11, using the photometric diameter of 2.0\arcsec\ is over-plotted with a green filled circle. From this we note that nine PNe from this sample (SMP\,S6, SMP\,S13, SMP\,S14, SMP\,S16, SMP\,S17, SMP\,S18, SMP\,S19, SMP\,S22 and SMP\,S24) follow the $S-\theta$ distribution of GPNe. The distinct, and possibly systematic, offset of the other five plotted objects in the region of high flux density in the GPN $S-\theta$ distribution is most likely caused by uncertainties in angular sizes and flux densities. However, it could also point to  different evolutionary pathways for PNe in low metallicity environments like MCs. More sensitive observations of a larger sample of SMC PNe are needed in order to examine this possibility.

Five objects (LIN\,45, SMP\,S11, LIN\,142, LIN\,321 and LIN\,339) show a large separation from the GPN $S-\theta$ distribution. Additionally, we include LIN\,41 in this group. In the various literature, we found that all of these objects have opposing views regarding their PN nature (see Section~\ref{sec:pnmimics}). Based on those previous results and our findings which include the MIR morphologies shown in (Fig.~\ref{fig:sageimages}) we conclude that none of these six objects are {\it bona fide} PNe.

\begin{figure*}[ht]
\centering
\includegraphics[trim=0 0 0 0, width=.3 \textwidth]{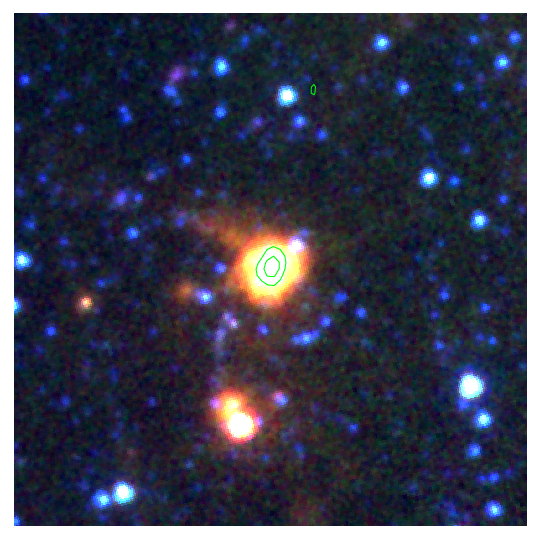}
\includegraphics[trim=0 0 0 0, width=.3\textwidth]{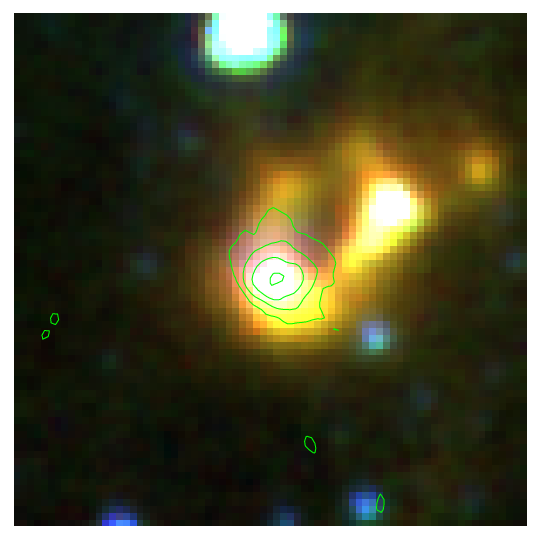}
\includegraphics[trim=0 0 0 0, width=.3\textwidth]{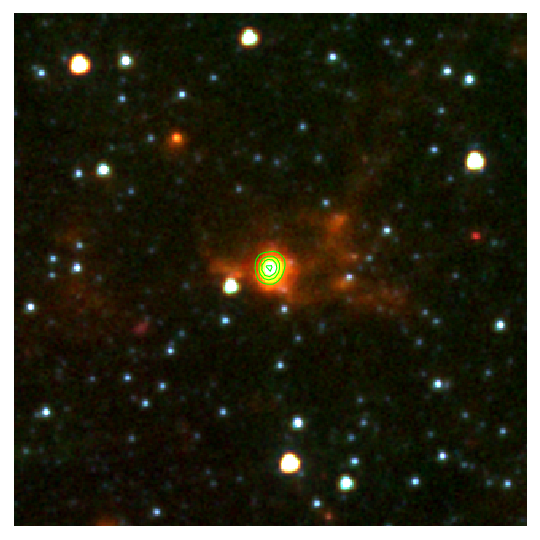}
\includegraphics[trim=0 0 0 0, width=.3\textwidth]{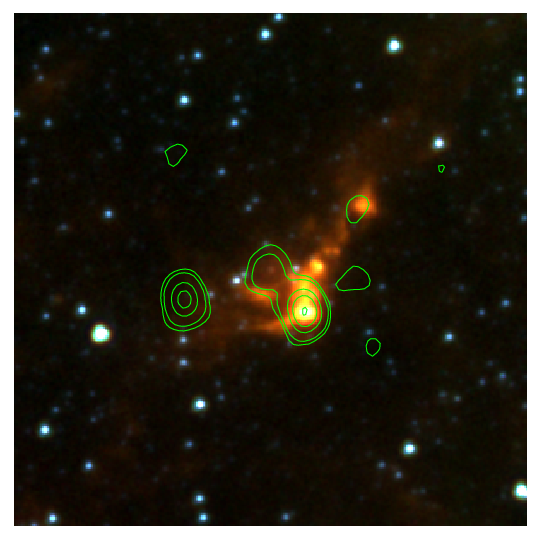}
\includegraphics[trim=0 0 0 0, width=.3\textwidth]{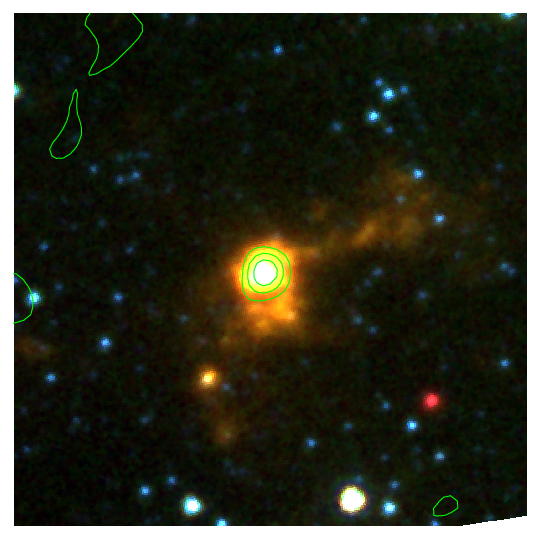}
\includegraphics[trim=0 0 0 0, width=.3 \textwidth]{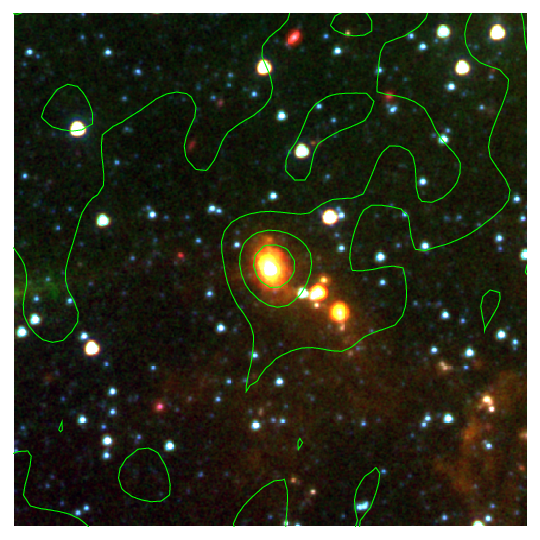}
\caption{Colour composite images from the IRAC (GLIMPSE-SMC) data with RGB = 8.0, 5.8 and 4.5 $\micron$ overlaid with radio contours. Top row (from left to right): LIN\,41, LIN\,45 and SMP\,S11. Bottom row (from left to right): LIN\,142, LIN\,321 and LIN\,339}
\label{fig:sageimages}
\end{figure*}

Finally, using the adopted 5~GHz flux densities and reported angular diameters we calculated electron densities ($n_e$) and ionised masses ($M_i$) for all positively detected objects (Table~\ref{tbl:physprop}, Cols.~8 and 9, respectively). The $n_e$ and $M_i$ were calculated from equations (1) and (2) in \cite{gathier1987}, respectively. We used the canonical electron temperature for all objects of $10^4$~K, solar metallicity parameters and filling factor of $0.5$ as an average between frequently used filling factors of 0.3 \citep{Boffi1994} and 0.75 \citep{gathier1987}. Using solar metallicity certainly adds a systematic error to the results but we estimate that that error is small compared to other sources of error. Uncertainties in $n_e$ were estimated to be on the  order of 40\%. The five objects with a doubtful PN nature show masses much larger than expected for PNe. The rest of the sample is in the range of ionised masses and electron densities for a population of relatively young PNe.

\section{SMC PN mimics}\label{sec:pnmimics}

{\it LIN\,41} was first identified as an emission star and an emission nebula in \cite{Henize1956}. \cite{Lindsay1961} proposed that LIN~41 is a possible planetary nebula. This view is disputed in \cite{Barlow1987,Monk1988,Oliveira2012} by suggesting that LIN~41 is very likely an \HII\ region powered by an early B-type main sequence star. In IRAC bands LIN~41 appears as a resolved source with colours typical for \HII\ regions and faint halo-like structures including a faint extension NE from the bright core. Based on the relatively low radio brightness (the radio flux density is 0.6~mJy) we propose that this object belongs to the population of ultra compact \HII\ regions.

{\it LIN\,45} was classified as a PN candidate by \cite{Lindsay1961} and as \HII\ region by \cite{Henize1963}. Spectral line ratios \citep{Dufour1977} and overall morphology in \Halpha\  \citep{McCall1990, Oey2013} suggest a compact \HII\ region with bright core or ridge, surrounding a late O or early B type star. However, \cite{Jacoby2002} noted that in the low metallicity environment of the SMC the \OIII\//\Halpha\ ratio of a PN would be similar to that of an \HII\ region and they re-considered LIN\,45 to be a PN. Finally, \cite{Hajduk2014} list it as a PN mimic with photometric variability. The MIR morphology, estimates of the ionised mass, and angular size at 6~cm clearly suggest that this object is unlikely a PN.

{\it SMP\,S11} is classified as a bipolar PN with a bright compact core and extended lobes \citep[up to 2\arcsec\ from the center:][]{Stanghellini2003}. \cite{Bernard-Salas2009} reported an unusual ``bump'' in the MIR continuum of this source with a peak of around 30-35~$\mu$m indicating cold dust. It showed the largest offset from the expected S abundance in a sample of 6 SMC PNe examined by \cite{k_Karakas_Ferland_Maguire_2012}. Finally, \cite{Shaw2010} noted that its metallicity is typical of an SMC \HII\ region and atypical of a bipolar PN suggesting that this object is not a PN. The separation from the GPN population in the $S-\theta$ diagram and its unusually high ionised mass can be attributed to uncertainties in the angular size and radio flux density. However, the MIR image shows complex structures around a bright core, much larger than for typical PNe. All of this suggests that this object is a PN mimic.

{\it LIN\,142} was originally classified as a PN by \cite{Henize1956}. It was later classified by \cite{DeOliveira2000} as a star cluster pair or multiple. \cite{Bica2008} classified it as a nebula, likely with an embedded cluster and as a member of a pair of stars. The separation from the GPN population in the $S-\theta$ diagram, its unusually high ionised mass, and large angular size suggest that this object is part of an \HII\ region (DEM\,S50) and not  a PN.

{\it LIN~321} and {\it LIN\,339} are listed as a PNe in \cite{Henize1956}. However, {\it LIN~321} was characterised as a Low Excitation Blob by \cite{Meynadier2007}.  {\it LIN~339} was classified as a compact \HII\ region by \cite{Charmandaris2008}. The separation from the GPN population in the $S-\theta$ diagram, their unusually high ionised masses, their large angular sizes, and  their morphologies in the MIR composite images suggest that both of these objects are very likely compact \HII\ regions.

\begin{figure}[t]
\includegraphics[width=0.9\columnwidth]{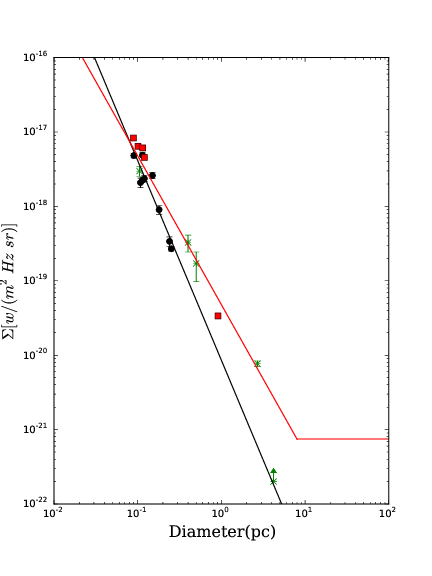}
\caption{The 6~cm radio surface brightness measurements $(\Sigma)$ of the  SMC PNe are shown here with round black markers with error bars. The square red markers are the five LMC PNe plotted by \cite{Vukotic2009} from data presented in \cite{Filipovic2009}. The red slanted line with the horizontal section at the bottom is the LMC 3$\sigma$ sensitivity plot for the LMC observations made by \cite{Filipovic2009}. The black slanted line is a fit of the SMC radio PN data to a power law function with an index of --2.7$\pm$0.4. The correlation coefficient is \mbox{--0.94.} The green x markers are the plots of the five PN mimics for which we have diameter data.  This figure re-plots a portion of Figure~1 from \cite{Vukotic2009} to include our SMC data. } %% no full stop at the end of caption
\label{fig:sigmaD}
\end{figure}

\section{$\Sigma- D$ relation for SMC PNe}

Surface brightness is measure independent of distance and has been applied to supernovae for over 50 years \citep{Shklovskii1960}. \cite{Amnuel1984} applied this measure to PNe and derived a basic $\Sigma-D$ relationship for PNe: $\Sigma\thicksim D^{-2}$. \cite{Urosevic2007} used 44 Galactic PN calibrators with distances of less than 0.7~kpc to determine the multiplier, A, and the exponent, $\beta$ in the equation $\Sigma=A \cdot D^{-\beta}$ ($\beta=2.07\pm0.19$) with a strong correlation coefficient (r) of 0.86.

\cite{Urosevic2009} used the interacting stellar winds model of \cite{Kwok1994} to derive a theoretical $\Sigma- D$ model for optically thin PNe. They noted that the majority of PNe will evolve from optically thick to optically thin but the time it takes for this to occur is dependent on many factors. In their model the fast wind from the central star sweeps up the slow wind from the previous evolutionary stage AGB star. This ``piling up'' creates the nebular shell. The optically thin PN model predicts a decreasing radio surface brightness from the enclosed volume as the nebular shell diameter increases. Early phases of evolution of the PN are shown to have a $\beta = 1$.  Later in the evolution of PNe the surface brightness, $\Sigma$, declines as $\beta$ tends towards a value of 3. Their empirical fits to various Galactic data sets from the literature with 5~GHz RC surface brightness and optically measured diameters result in values of $\beta$ from approximately 1.4 to 2.6. The data set with the best correlation coefficient \cite{Urosevic2009} used was Galactic data from PNe at a distance of $<1~kpc$ and found to be $\beta = 2.61 \pm 0.21$ with $r=-0.97$. With this distance limit there should be no Malmquist bias in their results. They suggest that the $\Sigma- D$ relation they derived should be used to indicate only general trends in surface brightness. Our SMC estimate of $\beta$ = $2.7\pm0.4$ with $r=-0.94$ is in excellent agreement with their Galactic results. Therefore, our SMC sample of PNe appear to be in a similar state of evolution.

We adopted the definition for the radio surface brightness employed by \cite{Vukotic2009} in order to make direct comparisons of our SMC data and their LMC data.

\begin{equation}\label{eqn:RCbrightness}
\Sigma \left[ {W{m^{ - 2}}H{z^{ - 1}}s{r^{ - 1}}} \right] = 1.505 \cdot {10^{ - 19}}{{S\left[ {Jy} \right]} \mathord{\left/
 {\vphantom {{S\left[ {Jy} \right]} {{\theta ^2}\left[ ' \right]}}} \right.
 \kern-\nulldelimiterspace} {{\theta ^2}\left[ ' \right]}}
\end{equation}

\noindent

We plotted our radio detected SMC PNe in Fig.~\ref{fig:sigmaD} modelled after the $\Sigma- D$ plot from \cite{Vukotic2009}. The black round points are the SMC PN measurements reported here which were used to calculate the power law index. The surface brightness sensitivity of our SMC measurements allow us to plot PNe at lower $\Sigma$ values than the LMC PNe plotted by \cite{Vukotic2009}. They found $\beta$ (for their fit to the 6~cm data) 0.9$\pm$0.5 with a correlation coefficient of 0.64. However, they suggested that the sensitivity selection effect and small number of data points above the sensitivity likely resulted in a value that ``should not be taken seriously''. The SMC PN data would also be expected to be under the influence of the sensitivity selection effect since only 17\% of the SMC PNe we examined were detectable at the sensitivity of our observations and virtually all PN are expected to be radio emitters until they disperse \citep{GurzadyanGrigorA.1997}. \cite{Vukotic2009} indicated that the flattening of $\beta$ due to a cutoff at lower flux densities is expected. From simulations by \cite{Vukotic2009}, however, for values of $\beta$ below 2 there should be little influence on the measured slope. For larger values of slope, the sensitivity effect should force the measured slope to be little more than 2. Our measured value of $\beta$ is expected to be subject to the flattening effect, however, we measure a $\beta$ = $2.7\pm0.4$ which is consistent with the Galactic data analysed by \cite{Urosevic2009}.

Our data set indicates the difficulty of culling true PNe from mimics. Fig.~\ref{fig:sigmaD} shows our data set and five objects which we have excluded from consideration. We note that two mimics, LIN\,45 and SMP\,S11, are very close to the surface brightness fit from the non-mimics in our study. All of these objects have been classified previously in the literature as PNe. Our examination indicates that they are actually PN mimics. The exclusion of the identified mimics results in the excellent agreement with the Galactic measurements that \cite{Urosevic2009} reported.

\section {Summary} \label{s:Summary}

We examined the positions of 105 catalogued SMC PNe that are listed in the SIMBAD database. These locations were analysed using all available radio maps at 3, 6, 13 and 20~cm. From these locations, sixteen sources were found to have detectable radio emissions in at least one of these radio bands with a flux density which is $\geq 3\sigma$ above the noise. Of these sixteen objects, ten have new radio detections reported here for the first time.

However, from the set of sixteen objects, we suggest that six of them are actually PN mimics. Two of our newly detected RC objects appear to be mimics as well as four previously detected RC objects.

Our measured $\Sigma- D$ relation for the SMC is consistent with similar measurements in the LMC and Milky Way.

In order to extend our understanding of the SMC and LMC PNe, we need more high quality measurements of the PNe in each galaxy. The improvements CABB brings to the ATCA observations of the SMC allowed us to more than double the population of known radio PNe in the SMC. The low noise and high-resolution characteristics of ATCA are well suited for this research and will lead to a much better understanding of extragalactic PNe, which will in turn further our understanding of Galactic PNe.

\section*{Acknowledgements}

The Australia Telescope Compact Array is part of the Australia Telescope National Facility which is funded by the Commonwealth of Australia for operation as a National Facility managed by CSIRO. This paper includes archived data obtained through the Australia Telescope Online Archive. The MCELS is funded through the support of the Dean B. McLaughlin fund at the University of Michigan and through NSF grant 9540747. This research has made use of Aladin, SIMBAD and VizieR, operated at the CDS, Strasbourg, France. We used the {\sc karma} and {\sc miriad} software packages developed by the ATNF and the {\sc python} programming language. We thank Dejan Uro\v sevi\'c and Branislav Vukoti\'c for their comments.  We would like to thank the anonymous referee for the astute comments and careful reading of this paper that significantly improved the final version.

\bibliographystyle{spr-mp-nameyear-cnd} %% BibTeX style
\bibliography{leverenz_bib}                %% BibTeX data

%\appendix

\end{document}